\documentclass[12pt]{iopart}
\usepackage{amssymb,graphicx,amsbsy}
\begin{document}

\title{Allometric Exponent and Randomness}
\author{Su Do Yi$^1$, Beom Jun Kim$^1$, and Petter Minnhagen$^2$}
\address{$^1$BK21 Physics Research Division and Department of Physics, Sungkyunkwan University, Suwon 440-746, Korea}
\address{$^2$IceLab, Department of Physics, Ume{\aa} University, 901 87 Ume{\aa}, Sweden}
\ead{beomjun@skku.edu  {\rm and}  petter.minnhagen@physics.umu.se}

\begin{abstract}
An allometric height-mass exponent $\gamma$ gives an approximative power-law
relation $\langle M\rangle \propto H^\gamma$ between the average mass $\langle M\rangle$
and the height $H$, for a sample of individuals. The individuals
in the present study are humans but could be any biological organism. The
sampling can be for a specific age of the individuals or for an age-interval.
The body-mass index (BMI) is often used for practical purposes
when characterizing humans and it is based on the allometric exponent $\gamma=2$.
It is here shown that the actual value of $\gamma$
is to large extent determined by the degree of correlation between mass and
height within the sample studied: no correlation between mass and height means
$\gamma=0$, whereas if there was a precise relation between mass and height
such that all individuals had the same shape and density then $\gamma=3$. The
connection is demonstrated by showing that the value of $\gamma$ can be
obtained directly from three numbers characterizing the spreads of the relevant
random Gaussian statistical distributions: the spread of the height and mass
distributions together with the spread of the mass distribution for the average
height. Possible implications for allometric relations in general are
discussed.

\end{abstract}


\maketitle

\section{Introduction}
Allometric relations in biology describe how a quantity $Y$ scales with the
body mass $M$, i.e., $Y=AM^{\frac{1}{\gamma}}$,  where $\gamma$ is an allometric exponent.
Allometric relations have a long history with pioneering work by D'Arcy Thomson
\emph{On Growth and Form}~\cite{thomson} and J.S. Huxley \emph{Problems of
Relative Growth}~\cite{huxley}. Among others, the allometric relation for the metabolic
rate $B$ has drawn much interest: Kleiber's law~\cite{kleiber} states that $B \sim M^p$,
with $p = 1/\gamma  \approx 3/4$, and has been tested in e.g. Refs.~\cite{dodds},\cite{white}, and \cite{simini}. For a review of allometric relations see Ref.~\cite{brown} \emph{Scaling in Biology}.
In the present study we focus on the allometric
relation between height and mass for humans. This mass ($M$) - height ($H$) relation has
an even longer history going back to the pioneering work by A. Quetelet \emph{A
Treatise on Man and the Developments of his Faculties} from
1842~\cite{quetelet}, where the allometric relation has been introduced to define
a normal man so that $M/H^\gamma$ becomes a Gaussian variable.
The precise definition of the allometric exponent used in the present study
is $\langle M \rangle_H \propto H^\gamma$ where $\langle M \rangle_H$ is the
average mass of the individuals of height $H$ in the sample. Note that the
allometric exponents $\gamma=1, 2$ and $3$ correspond to that mass is
proportional to height, body surface and body volume, respectively. For
practical purposes $\gamma=2$ is often a good approximation for humans, as
shown in Ref.~\cite{keys}. This approximation is the basis for the body mass index (BMI)
$A$ given by  $\langle M \rangle_H =AH^2$, provided mass is in kilogram and height in meter.
More recently it has been suggested that a larger allometric index $2< \gamma \leq 3$ should be
more appropriate~\cite{mackay,burton1,burton2}. In particular Burton in Ref.~\cite{burton1} suggests that $\gamma=2$ is an underestimate caused by randomness. This is in accordance with the conclusions reached in the present investigation.

The object with the present investigation is to understand the relation between
the exponent $\gamma$ and the randomness for a given sample of individuals. The
issue is best illustrated by a specific example. Figure~\ref{fig:prb}(a) and (b) show the
height and mass distributions, $P(H)$ and $P(M)$, respectively,
for 25000 children 18 years old from Hong
Kong~\cite{hongkong}. Figure~\ref{fig:prb}(c) in addition shows the distribution
$P(M|H = \langle H\rangle)$ for the children of average height $\langle H\rangle$.
All these three statistical
distributions are to very good approximation Gaussians. This means that the
variables in all three cases are randomly distributed around their respective
average values. The random spread are in all three cases characterized be the
normalized standard deviations, which we denote $\tilde{\sigma}_H$,
$\tilde{\sigma}_M$, and $\tilde{\sigma}$ for random spread of height, mass and
mass-for-average-height, respectively. The relation derived in the present
paper states that $\gamma$ to good approximation should be given by
$\gamma=\frac{\sqrt{\tilde{\sigma}_M^2-\tilde{\sigma}^2}}{\tilde{\sigma_H}}$.
From the random spreads in Fig.~\ref{fig:prb} one  then
finds $\gamma=1.63$. Figure~\ref{fig:MH}(a) shows that this is a very accurate
prediction. This means that the allometric exponent is entirely determined by the
randomness of the three distributions. Why is this so and what does it imply?
These are questions which come to mind.

In Sec.~\ref{sec:relation} the relation between $\gamma$ and the random spreads
is derived. Comparisons with data are made in Sec.~\ref{sec:comp}, whereas we in
Sec.~\ref{sec:sum} sum up and discuss the results.

\section{Allometric exponent expressed in normalized standard deviations}
\label{sec:relation}
The point made in the present paper is that the exponent $\gamma$ can be estimated
from the sole knowledge of the first and second moments of the mass  and
height distributions. In order to derive such a relation we assume that the
mass and height distributions are approximately Gaussians. This is, as is
illustrated by the datasets in Fig.~\ref{fig:prb},
often a fair approximation around the maxima of the distributions. It means
that the probability distribution for the mass and height distributions are
approximately given by

\begin{equation}
\label{eq:P_M}
P_M(M)=\frac{1}{\sqrt{2\pi\sigma_M^2}}\exp\left(-\frac{(M-\langle M\rangle)^2}{2\sigma_M^2}\right),
\end{equation}

\begin{equation}
\label{eq:P_H}
P_H(H)=\frac{1}{\sqrt{2\pi\sigma_H^2}}\exp\left(-\frac{(H-\langle H\rangle)^2}{2\sigma_H^2}\right),
\end{equation}
respectively. Note that these two distributions are characterized by the four
explicit numbers $\langle M\rangle$, $\langle H\rangle$, $\sigma_M$ and
$\sigma_H$. The degree of correlation between the mass and height is then given
by the mass distribution for a given height, which we likewise assume to be
approximately Gaussian and is given by the conditional probability
\begin{equation}
\label{eq:M|H}
P(M|H)=\frac{1}{\sqrt{2\pi\sigma(H)^2}}\exp\left(-\frac{(M-\langle M\rangle_H)^2}{2\sigma(H)^2}\right),
\end{equation}
where $\langle M \rangle_H$ and $\sigma(H)$ are the average and the standard deviation of mass
obtained for all individuals of height $H$.  Note that the standard deviation $\sigma_X$ for a stochastic variable $X$ is related to the first and second moments by $\sigma_X^2=\langle X^2\rangle -\langle X \rangle^2$.
A particular feature in the present context is that the distribution of mass
for a given height can approximately be characterized by a constant standard
deviation $\sigma$, since in practice it turns out that $\sigma(H)$ is only
weakly dependent on $H$ in the close vicinity of $\langle H\rangle$ [see
Fig.~\ref{fig:prb}(d)].
Thus Eq.~(\ref{eq:M|H}) can approximately be reduced to
\begin{equation}
\label{eq:M|H1}
P(M|H)=\frac{1}{\sqrt{2\pi\sigma^2}}\exp\left(-\frac{(M-\langle M\rangle_H)^2}{2\sigma^2}\right) .
\end{equation}
Another particular feature of the mass relation is that the average mass for a given height, $\langle M\rangle_H$, monotonously increases with height. We can use this one-to-one correspondence by changing the variable in Eq.~(\ref{eq:P_H}) from $H$ to $\langle M\rangle_H$. The height distribution in terms of
$\langle M\rangle_H$ is then just $ P_H(H(\langle M\rangle_H))$. This means
that there exists a precise relation between the three distributions
[Eqs.~(\ref{eq:P_M}), (\ref{eq:P_H}), and  (\ref{eq:M|H1})] given by
\begin{equation}
\label{eq:conv}
P_M(M)=\int d\langle M\rangle_H P(M|\langle M\rangle_H)P_H(H(\langle M\rangle_H))\frac{dH(\langle M\rangle_H)}{d\langle M\rangle_H} .
\end{equation}

Another crucial feature of the data is that $\langle M\rangle_H$ to some approximation is described by the power-law relation\begin{equation}
\label{eq:bmi}
\langle M\rangle_H =AH^\gamma .
\end{equation}
This is the allometric relation in focus and discussed in the present paper.
Here $A$ and $\gamma$ are two constants. The constant $A$ can be expressed as
$A=\langle M\rangle_H H^{-\gamma} =
\langle M\rangle_{\langle H\rangle} \langle H\rangle^{-\gamma}$. However, for Gaussian distributions
$\langle H\rangle$ corresponds to the peak position of the height distribution
and, since the individuals in this peak also to good approximation have the
average mass, it follows that $\langle M\rangle_{\langle H\rangle} \approx \langle M\rangle$.
In the following, we will consequently use the simplified
estimate $A=\langle M\rangle \langle H\rangle^{-\gamma}$. The argument for
$P_H$ in Eq.~(\ref{eq:P_H}) is
\begin{equation}
\label{eq:H(M)}
H(\langle M\rangle_H)=\left(\frac{\langle M\rangle_H}{A}\right)^{\frac{1}{\gamma}}=
\langle H\rangle \left(\frac{\langle M\rangle_H}{\langle M\rangle}\right)^{\frac{1}{\gamma}} .
\end{equation}
Close to the peak of this distribution at $\langle M_{\langle H\rangle} \rangle$ we can use the linear approximation
\begin{equation}
\label{eq:lin}
H(\langle M\rangle_H)=\langle H\rangle \left(\frac{\langle M\rangle_H}{\langle M\rangle}\right)^{\frac{1}{\gamma}}
\approx \langle H\rangle \left(1+\frac{1}{\gamma\langle M\rangle}(\langle M\rangle_H-\langle M\rangle)\right) .
\end{equation}
Inserting this approximation into the relation \[P_H(\langle (M)\rangle)=P_H(H(\langle M\rangle_H))\frac{dH(\langle
M\rangle_H)}{d\langle M\rangle_H}\] leads to the Gaussian distribution
\begin{equation}
\label{eq:P_HM}
P_H(\langle (M)\rangle =\frac{1}{\sqrt{2\pi\sigma_H^2\gamma^2\langle M\rangle^2\langle H\rangle^{-2}}}\exp\left(-\frac{(\langle
M\rangle_H-\langle M_{\langle H\rangle} \rangle)^2}{2\sigma_H^2\gamma^2\langle M\rangle^2\langle H\rangle^{-2}}\right) .
\end{equation}
Using Eq.~(\ref{eq:P_HM}) together with Eq.~(\ref{eq:M|H1}), means that the right-hand side of  Eq.~(\ref{eq:conv}) becomes a convolution of two Gaussian. Since the convolution of two Gaussians with standard deviations $\sigma_1$ and $\sigma_2$ becomes a Gaussian with standard deviation $\sigma_3=\sqrt{\sigma_1^2+ \sigma_2^2}$, it follows that $\sigma_M^2=\sigma^2+\sigma_H^2\gamma^2\langle H\rangle^2/\langle M\rangle^2$ or equivalently

\begin{equation}
\label{eq:gamma}
\gamma=\frac{\langle H\rangle}{\langle M \rangle}\frac{\sqrt{\sigma_M^2-\sigma^2}}{\sigma_H}=\frac{\sqrt{\tilde{\sigma}_M^2-\tilde{\sigma}^2}}{\tilde{\sigma_H}} ,
\end{equation}
where we have introduced the normalized standard deviations
$\tilde{\sigma}_M=\sigma_M/\langle M\rangle$,$\tilde{\sigma}=\sigma/\langle
M\rangle$ and $\tilde{\sigma}_H=\sigma_H/\langle H\rangle$.
Equation~(\ref{eq:gamma}) is the central relation in the present investigation and shows that $\gamma$ can be approximately obtained from the three dimensionless numbers $\tilde{\sigma}_M$, $\tilde{\sigma}$ and $\tilde{\sigma}_H$, which measures the random spread of the data in units of, respectively, the average mass and height of the individuals.

Also note that $\gamma$ given by Eq.~(\ref{eq:gamma}) is what you get when an allometric relation is used as an \emph{ansatz}. It does not \emph{a priori} say anything about whether or not an allometric relation is a good approximation of the data.

\section{Comparison with data}
\label{sec:comp}

\begin{figure}
\begin{center}
\includegraphics[width=0.9\textwidth]{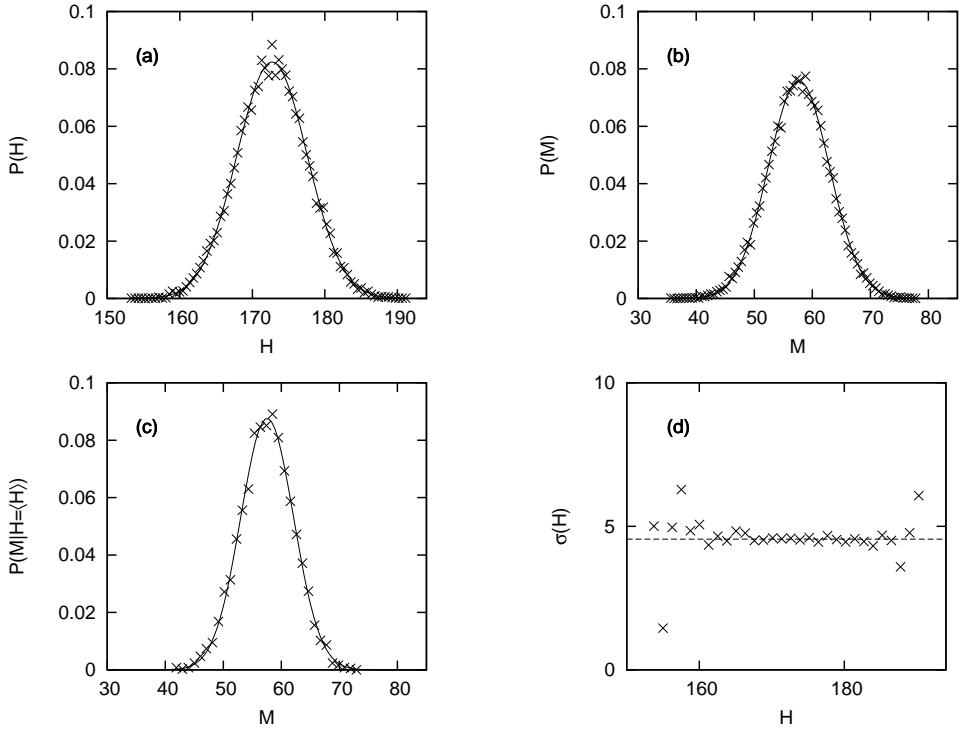}
\end{center}
\caption{(a) The height distribution $P(H)$ and (b) the mass distribution $P(M)$
for 25000 Hong Kong children 18 years old~\cite{hongkong}.
(c) The conditional probability distribution
$P(M|H = \langle H \rangle)$ obtained for 3670 children whose heights are in the interval
$[171.82, 173.62]$ around $\langle H \rangle = 172.72$.
In (a)-(c), the crosses are the data and the full-drawn curves are the
corresponding Gaussian approximations.
The numbers of bins are 81 for (a) and (b), and 31 for (c).
$H$ in (a) and $M$ in (b) and (c) are in units of cm and kg.
(d) The standard deviation $\sigma(H)$ of the distribution $P(M|H)$ in Eq.~(\ref{eq:M|H}) as a function of height $H$.
The horizontal line shows that the standard deviation $\sigma(H)$ is independent of $H$ in a range
around the average height $\langle H \rangle \approx 172.72$ cm.

}
\label{fig:prb}
\end{figure}

\begin{figure}
\begin{center}
\includegraphics[width=0.9\textwidth]{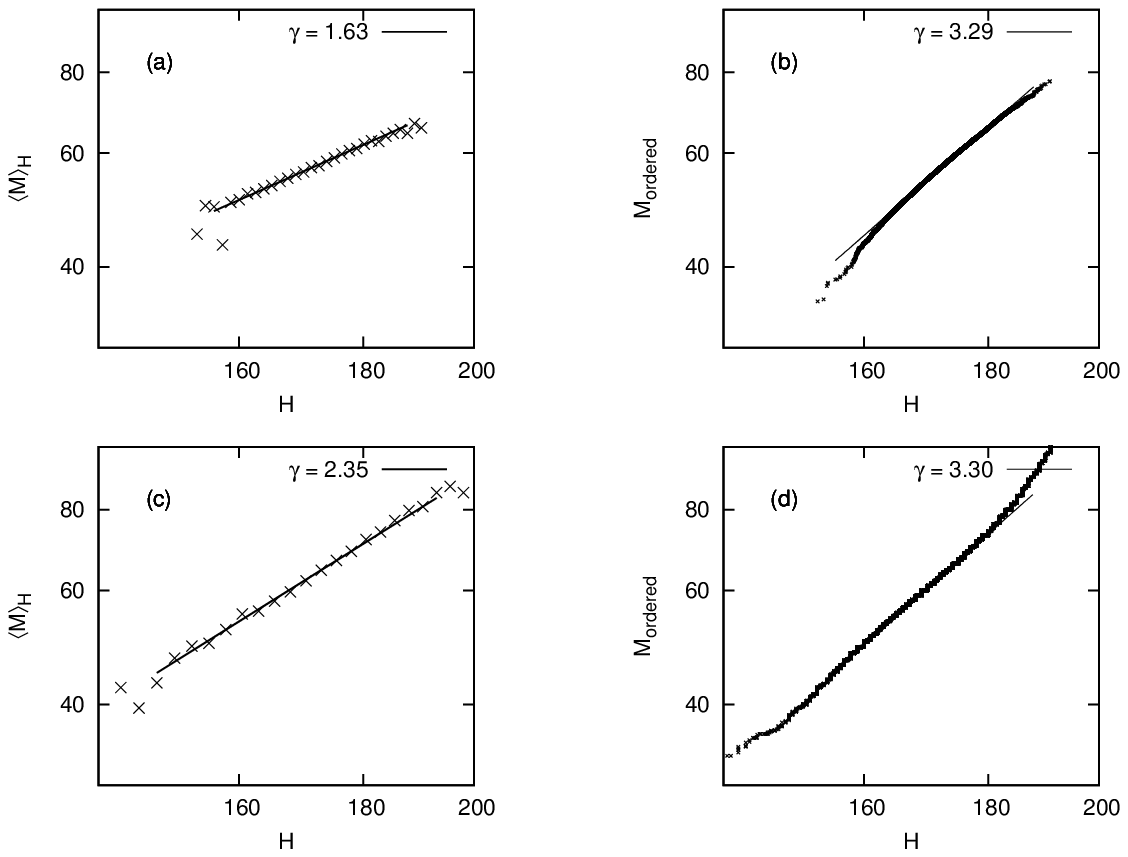}
\end{center}
\caption{Allometric relations for the Hong Kong data in (a) and (b) and for the Swedish data in (c) and (d):
 (a) Log-log plot of the average mass $\langle M\rangle_H$ as a
function of height $H$. Symbols correspond to the average for a length
interval of $1.26$ cm. The data fall on a straight line in accordance with the
allometric relation $\langle M\rangle_H\propto H^\gamma$. The value of $\gamma$
determined by the least-square fit to the data is $\gamma_{ex}=1.63$. The
straight line is the prediction in Eq.~(\ref{eq:explicit})
in terms of the random spreads given by
Eq.~(\ref{eq:gamma}). The prediction is very accurate for this dataset.
(b) Log-log plot of the ordered data $M_{\rm ordered}(H)$ as a function of $H$.
The data is well represented by $M_{\rm ordered}\propto H^\gamma$
with $\gamma_{\rm ordered}=3.29$.
(c) and (d) is the same as (a) and (b) for the Swedish data. The straight lines are least square fits to the data giving, respectively $\gamma_{ex}=2.35$ and $\gamma_{\rm ordered}=3.30$. These predictions are again in good agreement with the predictions given in Table~\ref{tab:gamma}.
Note that the allometric exponent between the Hong Kong data set in (a) and the Swedish data set in (c) are significantly different, whereas they are almost identical for within the ordered data-representation given by (b) and (d). These features are explained in the present paper.
}
\label{fig:MH}
\end{figure}

 In the light of the above theoretical underpinning we return to the data for
25000 children 18 years~\cite{hongkong}. One notes in Fig.~\ref{fig:prb} that both the
height  and the mass distributions to good approximations are Gaussians for
this dataset. The average mass for a child is $\langle M\rangle \approx 57.7$
kg and height $\langle H\rangle \approx 172.7$ cm. The
standard deviations are $\sigma_M \approx 5.3$ kg and $\sigma_H\approx 4.8$ cm. Figure~\ref{fig:prb}(c) shows that also the distribution of mass for given heights are
Gaussians and Fig.~\ref{fig:prb}(d) shows that the standard deviation $\sigma(H)$ in
Eq.~(\ref{eq:M|H}) is constant in a broad rage of $H$ around $\langle H \rangle$,
so that Eq.~(\ref{eq:M|H1}) gives a very good
description. As shown in Sec.~\ref{sec:relation}, under these conditions the relation given
by Eq.~(\ref{eq:gamma}) applies. This relation states that if there is a power-law
relation between average mass and height, $\langle M\rangle_H=AH^\gamma$,
then the best prediction for the given information is
 \begin{equation}
 \label{eq:explicit}
 \langle M\rangle_H=\langle M\rangle \left(\frac{H}{\langle H\rangle}\right)^{\frac{\sqrt{\tilde{\sigma}_M^2-\tilde{\sigma}^2}}{\tilde{\sigma}_H}}
 \end{equation}

 \begin{table}[t]
	\centering		
\begin{tabular}{c|c|c|c|c|c|c|c|c|c}       \hline
 & $\langle H \rangle $  & $\langle M \rangle $ & $\tilde{\sigma}_H $  &  $\tilde{\sigma}_M$ & $\tilde{\sigma}$ &
$\gamma_{th}$ &$\gamma_{ex}$ & $\gamma_{{\sigma=0}}$ & $\gamma_{\rm ordered} $\\ \hline
Hong Kong &  172.72    & 57.68  &   0.0280    &    0.0912 &  0.079 & 1.63 & 1.63 & 3.26 & 3.29 \\ \hline
Sweden &  170.48    & 60.52  &   0.0573    &    0.1862 &   0.130   & 2.33  & 2.35 & 3.25 & 3.30 \\ \hline
\end{tabular}	
\caption{Summary of 25000 data for children 18 years old from Hong Kong~\cite{hongkong} (first row) and for 11300 data for Swedish children in the year interval 13.5-19 years old~\cite{werner} (second row).
The average height $\langle H\rangle$ and the average mass $\langle M\rangle$
are in units of cm and kg. The normalized dimensionless
standard deviations for the height $\tilde{\sigma}_H$, the mass $\tilde{\sigma}_M$,
and for the mass distribution at average height $\tilde{\sigma}$ are listed.
The theoretical prediction $\gamma_{th}$  from Eq.~(\ref{eq:gamma}) and $\gamma_{ex}$, obtained
from the least-square fit to the data presented in Figs.~\ref{fig:MH}(a) and (c), are in good
agreement.  $\gamma_{\sigma=0}$ from Eq.~(\ref{eq:gamma}) with $\sigma = 0$
and $\gamma_{\rm ordered}$, obtained from the least-square fit to the data in
Figs.~\ref{fig:MH}(b) and (d), also agree with each other.
Note the close agreements between fitted and predicted
values of the allometric exponents $\gamma$ in all cases}
\label{tab:gamma}
\end{table}

\begin{table}[t]
\begin{small}
\centering		
\begin{tabular}{c||c|c|c|c|c}       \hline
&  $H$ vs $M$ & $H$ vs $M/H^3$ & $H$ vs $M/H^2$ & $H$ vs $M/H^{\gamma_{th}}$  & $H$ vs $M/H^{\gamma_{ex}}$ \\ \hline
Pearson's $r$ (Hong Kong) & 0.50   &  $-0.43$ & $-0.12$ & 0.01     & 0.01   \\ \hline
Pearson's $r$ (Sweden) & 0.64   &  $-0.21$ & $0.14$ &   0.03   & 0.02   \\ \hline \hline
\end{tabular}	
\caption{Pearson's correlation coefficient $r$ between the height $H$ and various quantities
($M$, $M/H^3$, $M/H^2$, $M/H^{\gamma_{th}}$, and $M/H^{\gamma_{ex}}$)
are computed for Hong Kong children~\cite{hongkong} (first row) and the Swedish data~\cite{werner} (second row).
The height-mass correlation ($r$ for $H$ vs $M$) is of course positively significant in both cases. For the Hong Kong data in the first row
the correlations between $H$ and $M/H^3$, and between $H$ and $M/H^2$ are negative,
implying that the exponents $2$ and $3$ are overestimations.
In contrast, $M/H^{\gamma_{ex}}$ and $M/H^{\gamma_{th}}$ exhibit neutral correlation with $H$,
which shows that our estimation $\gamma_{th} \approx 1.63$ describes data much better than
the conventional BMI-value $\gamma = 2$. Likewise in the second row for the Swedish data the correlations between $H$ and $M/H^3$, and between $H$ and $M/H^2$ are, respectively, negative and positive, implying an exponent between 2 and 3. This is again in agreement with our analysis and theory.
}
\label{tab:pearson}
\end{small}
\end{table}

Table~\ref{tab:gamma} gives the average height and mass (in cm and kg,
respectively) together with the three normalized standard deviations for the
statistical distributions: $\tilde{\sigma}_H $, $\tilde{\sigma}_M$ and
$\tilde{\sigma}$. The resulting power-law exponent predicted by
$\gamma_{th}=\frac{\sqrt{\tilde{\sigma}_M^2-\tilde{\sigma}^2}}{\tilde{\sigma_H}}$,
as well as $\gamma_{ex}$ obtained by direct fitting to the data
[see Fig.~\ref{fig:MH}(a)], is also listed.
The agreement between $\gamma_{th}$ and $\gamma_{ex}$ is very
precise and confirms that there really exists a relation between the spreads
and the power-law exponent. The question is what it implies.

In order to get an idea of what this means we note that if $\sigma=0$ then
there exists a one-to-one function between $M$ and $H$ and according to
Eq.~(\ref{eq:gamma}), we get
\begin{equation}
\label{eq:gamma0}
\gamma=\frac{\tilde{\sigma}_M}{\tilde{\sigma}_H} .
\end{equation}
Changing $\tilde{\sigma}$ in the Hong Kong children data to $\tilde{\sigma}=0$ changes
the prediction for $\gamma_{th}$ from 1.63 to 3.26 (compare
Table~\ref{tab:gamma} and Fig.~\ref{fig:gamma}). We can test this prediction
against the children data by re-ordering so that the children are assigned
masses which strictly follow the heights of the children. For this re-ordered
data, $\sigma$ is indeed zero and as shown in Fig.~\ref{fig:MH}(b) the slope
for this re-ordered data is indeed again close to the prediction. This gives a
direct demonstration of the connection between spread and power-law exponent.

Table~\ref{tab:pearson} gives various Pearson's $r$-coefficients computed
for various pairs of quantities. For the Hong Kong children data (first row of the table)
the height-mass correlation ($H$ vs $M$) is significantly positive,
implying that in general the taller child has the heavier mass.
It is to be noted that the correlations for $M/H^3$ and $M/H^2$ deviate much from zero, while
$M/H^{\gamma_{ex}}$ and $M/H^{\gamma_{th}}$ exhibit neutral correlation with the height,
suggesting that our estimation $\gamma_{th} \approx 1.63$ describes data much better than
the conventionally used BMI value $\gamma = 2$.

In order to rule out that there is anything accidental or fortuitous about the results presented, we have investigated a second data-set in the same way. This second data set gives height and mass for Swedish children between 13.5 to 19 years old (more precisely between 5000 to 7000 days old containing in total 11327 data points)~\cite{werner}. The results are presented in Fig.~\ref{fig:MH}(c) and (d) with parameters given in the second rows of Tables~\ref{tab:gamma} and ~\ref{tab:pearson}. From Table~\ref{tab:gamma} one can see that the average height and mass for these two data-sets are roughly the same. However, since the Swedish children data spans over a longer age period than the Hong Kong data, the standard deviations for height, $\tilde{\sigma}_H $, and mass, $\tilde{\sigma}_M$, are larger by about a factor 2. This is of course because children during a longer period grows more. Yet the ratio between the standard deviations, $\tilde{\sigma}_M/\tilde{\sigma}_H$, is closely equal for the two data-sets (3.26 and 3.25 respectively.) This ratio is in fact the $\gamma_{{\sigma=0}}$ and, as seen from Table~\ref{tab:gamma} and Figs.~\ref{fig:MH}(b) and (d), $\gamma_{{\sigma=0}}$ gives very precise estimates of the allometric exponents $\gamma_{\rm ordered} $. The fact that the exponents $\gamma_{\rm ordered} $ are very nearly the same for the two data-sets, suggests that, in this particular aspect, children from Hong Kong and Sweden are very similar. Also for the Swedish data set there is a good agreement between the experimental allometric exponent
$\gamma_{ex}$ and the prediction $\gamma_{th}$ from Eq.~(\ref{eq:gamma}) (compare Figs.~\ref{fig:MH}(c) and Table~\ref{tab:gamma}). However, there is a significant difference between the allometric exponents $\gamma_{ex}$ for the two data-sets: $\gamma_{ex}=1,63$ and 2.35 for, respectively, Hong Kong and Swedish children. The close agreement between $\gamma_{ex}$ and the prediction $\gamma_{th}$ for both data-sets suggests that the difference in value of the allometric exponent $\gamma_{ex}$ can be attributed to a relatively larger spread in weight for the children of average height for Hong Kong children compared to Swedish children. From this point of view it is rather a sampling difference than some difference in trait of a Hong Kong and Swedish individual. A possible explanation could be that, since Hong Kong  compared to Sweden for a long time has been a human hotspot with influx of people of great variety in both genetical and cultural backgrounds, this has resulted in a relatively larger spread of weight for a given height in a particular age interval.

\section{Discussion}
\label{sec:sum}
The implication of theses results become clearer when comparing
Fig.~\ref{fig:MH}(a) and Fig.~\ref{fig:MH}(b). Both represent data with the
same two Gaussian distributions for mass and height given in
Fig.~\ref{fig:prb}. The difference is that the data in Fig.~\ref{fig:MH}(a) also
has a spread of mass for individuals with a given height, as shown in
Fig.~\ref{fig:prb}. For the artificial data in Fig.~\ref{fig:MH}(b) there is
no such spread. The data in Fig.~\ref{fig:MH}(b) represents a true allometric
relation between mass and height of the form $M\propto H^\gamma$: as soon as
you pick a person with a certain height you also to very good approximation
know his mass. Since $\gamma$ for this artificial dataset is 3.29, it either
means that these artificial people either all have the same shape but get a bit
denser with increasing height, or that they just become somewhat
disproportionally fatter with increasing height. The point is that you in this
case can relate the allometric exponent to some property of the individual.
However, for the real data in Fig.~\ref{fig:MH}(a) this becomes more problematic.
This is because for a given height the individuals have a random mass
distributed about the mean, as shown in Fig.~\ref{fig:prb}(c). This random
spread can have a multitude of different causes, like availability of food,
climate, diseases, genetics etc. Different individuals are affected by these
multitude of causes in different ways.

An alternative way of describing this is as follows: Suppose you have a dataset like the one discussed in the present paper and suppose that each \emph{individual} can be characterized by the allometric relation $M\propto \langle H\rangle_M^{\gamma_0}$,
where $\langle H\rangle_M$ is the average height for individuals of mass $M$.
This means that if you pick an individual with mass $M$ then you know that his most likely height is given by allometric relation with $\gamma_0$. If in addition there was no randomness in the height, then you for certain know his height and the result is given by Fig.~\ref{fig:MH}(b). But if there is an randomness in the height caused by many causes, so that the probability of the height for an individual is described by a Gaussian probability distribution, then the allometric exponent becomes smaller and you can end up with something like Fig.~\ref{fig:MH}(a) instead. The difference is that
Fig.~\ref{fig:MH}(a) describes the {\em collective} dataset, whereas Fig.~\ref{fig:MH}(b) corresponds to an allometric relation on an {\em individual} basis.

\begin{figure}
\begin{center}
\includegraphics[width=0.9\textwidth]{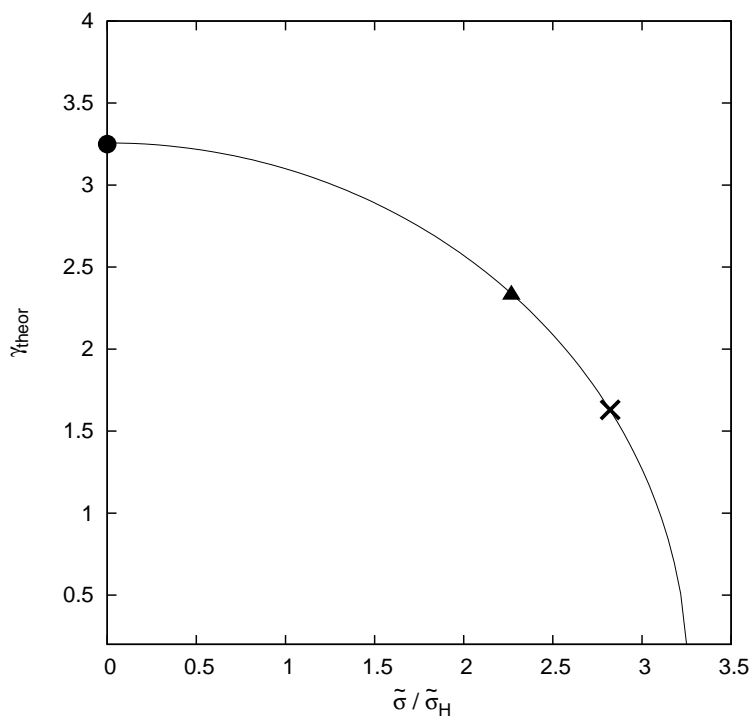}
\end{center}
\caption{Plot of the $\gamma_{th}$ obtained from Eq.~(\ref{eq:gamma}). The standard deviations $\tilde{\sigma}_M$ and $\tilde{\sigma}_H$ from the children data are used to plot $\gamma_{th}$ as a function of the standard deviation ratio $\sigma/\tilde{\sigma}_H$. Note that, since $\gamma_{\sigma=0}=\tilde{\sigma}_M/{\tilde{\sigma}_H}$ is almost identical for the Hong Kong and Swedish data, both predictions are obtained from the same curve. The black dot represent the predicted maximum value $\gamma_{{\sigma=0}}$ for $\sigma=0$. The cross and triangle represents the prediction of $\gamma_{th}$ for, respectively, the Hong Kong and Swedish children data.}
\label{fig:gamma}
\end{figure}

In Fig.~\ref{fig:MH} (a) and (b) it is precisely the random spread
which causes the decrease of the allometric exponent from 3.29 to 1.63. So you
cannot any longer associate the allometric exponent $\gamma$ with some unique growth
property of the individuals. It is rather like that the spread in
Fig.~\ref{fig:prb}(c) just tells you how much the masses and heights for
individuals are random and uncorrelated. This is also reflected by the
prediction given by Eq.~(\ref{eq:gamma}) which decreases from its maximum value
$\gamma=\frac{\tilde{\sigma}_M}{\tilde{\sigma}_H}$ to zero with increasing
random spread $\tilde{\sigma}$, as illustrated in Fig.~\ref{fig:gamma}.

Some further insight to this is given by the comparison between the Hong Kong children and the Swedish children: The allometric exponent for the Hong Kong children is significantly smaller than for the Swedish children. According to the present analysis, this difference can be traced to the difference in the relatively larger  spread in mass for a given height in case of the
Hong Kong data-set. One possible explanation for this difference in spread could be a sampling difference: Compared to Sweden,
Hong Kong has been a historical hotspot leading to a greater variety of people both from a genetical and a cultural point of view. Such greater variety is also likely to cause a larger variety in weight for a given height.

The present analysis is quite general and its implication is likely to have a wider range of applicability within allometric relations than just to the illustrative example of mass-height relation for humans, discussed here: It brings a caution against attributing too much specific cause to the precise value of an allometric exponent. The crucial point is that the allometric exponent for an individual is, because of randomness, not the same as the allometric exponent of the collective dataset.

\ack
This research was supported by Korea-Sweden Research Cooperation Program
through the National Research Foundation of Korea (NRF)
funded by the Ministry of Education, Science and Technology (2011-0031307).

\section*{References}

\end{document}